\documentclass{svmult}
\usepackage{epsfig,graphicx} 
\usepackage[bottom]{footmisc}
\usepackage{natbib,url}      
\bibpunct{(}{)}{;}{a}{}{,}   
\def\cite#1{\citealp{#1}}    
\def\authorindex#1{}  
\def\figspath{.}      

\begin{document}\newcount\preprintheader\preprintheader=1



\def\thisvolume{these proceedings}

\def\aj{{AJ}}			
\def\araa{{ARA\&A}}		
\def\apj{{ApJ}}			
\def\apjl{{ApJ}}		
\def\apjs{{ApJS}}		
\def\ao{{Appl.\ Optics}} 
\def\apss{{Ap\&SS}}		
\def\aap{{A\&A}}		
\def\aapr{{A\&A~Rev.}}		
\def\aaps{{A\&AS}}		
\def\an{{Astron.\ Nachrichten}}
\def\aspcs{{ASP Conf.\ Ser.}}
\def\assp{{Astrophys.\ \& Space Sci.\ Procs., Springer, Heidelberg}}
\def\azh{{AZh}}			
\def\baas{{BAAS}}		
\def\jrasc{{JRASC}}	
\def\memras{{MmRAS}}		
\def\mnras{{MNRAS}}
\def\nat{{Nat}}		
\def\pra{{Phys.\ Rev.\ A}} 
\def\prb{{Phys.\ Rev.\ B}}		
\def\prc{{Phys.\ Rev.\ C}}		
\def\prd{{Phys.\ Rev.\ D}}		
\def\prl{{Phys.\ Rev.\ Lett.}} 
\def\pasp{{PASP}}
\def\pasj{{PASJ}}		
\def\qjras{{QJRAS}}
\def\science{{Sci}}		
\def\skytel{{S\&T}}		
\def\solphys{{Solar\ Phys.}} 
\def\sovast{{Soviet\ Ast.}}  
\def\ssr{{Space\ Sci.\ Rev.}}
\def\svassp{{Astrophys.\ Space Sci.\ Procs., Springer, Heidelberg}}
\def\zap{{ZAp}}			
\let\astap=\aap
\let\apjlett=\apjl
\let\apjsupp=\apjs
\def\grl{{Geophys.\ Res.\ Lett.}}  
\def\jgr{{J. Geophys.\ Res.}} 

\def\ion#1#2{{\rm #1}\,{\uppercase{#2}}}  
\def\deg{\hbox{$^\circ$}}
\def\sun{\hbox{$\odot$}}
\def\earth{\hbox{$\oplus$}}
\def\la{\mathrel{\hbox{\rlap{\hbox{\lower4pt\hbox{$\sim$}}}\hbox{$<$}}}}
\def\ga{\mathrel{\hbox{\rlap{\hbox{\lower4pt\hbox{$\sim$}}}\hbox{$>$}}}}
\def\sq{\hbox{\rlap{$\sqcap$}$\sqcup$}}
\def\arcmin{\hbox{$^\prime$}}
\def\arcsec{\hbox{$^{\prime\prime}$}}
\def\fd{\hbox{$.\!\!^{\rm d}$}}
\def\fh{\hbox{$.\!\!^{\rm h}$}}
\def\fm{\hbox{$.\!\!^{\rm m}$}}
\def\fs{\hbox{$.\!\!^{\rm s}$}}
\def\fdg{\hbox{$.\!\!^\circ$}}
\def\farcm{\hbox{$.\mkern-4mu^\prime$}}
\def\farcs{\hbox{$.\!\!^{\prime\prime}$}}
\def\fp{\hbox{$.\!\!^{\scriptscriptstyle\rm p}$}}
\def\micron{\hbox{$\mu$m}}
\def\onehalf{\hbox{$\,^1\!/_2$}}	
\def\onethird{\hbox{$\,^1\!/_3$}}
\def\twothirds{\hbox{$\,^2\!/_3$}}
\def\onequarter{\hbox{$\,^1\!/_4$}}
\def\threequarters{\hbox{$\,^3\!/_4$}}
\def\ubv{\hbox{$U\!BV$}}		
\def\ubvr{\hbox{$U\!BV\!R$}}		
\def\ubvri{\hbox{$U\!BV\!RI$}}		
\def\ubvrij{\hbox{$U\!BV\!RI\!J$}}		
\def\ubvrijh{\hbox{$U\!BV\!RI\!J\!H$}}		
\def\ubvrijhk{\hbox{$U\!BV\!RI\!J\!H\!K$}}		
\def\ub{\hbox{$U\!-\!B$}}		
\def\bv{\hbox{$B\!-\!V$}}		
\def\vr{\hbox{$V\!-\!R$}}		
\def\ur{\hbox{$U\!-\!R$}}


\def\labelitemi{{\bf --}}  

\def\rmit#1{{\it #1}}              
\def\rmit#1{{\rm #1}}              
\def\etal{\rmit{et al.}}           
\def\etc{\rmit{etc.}}           
\def\ie{\rmit{i.e.,}}              
\def\eg{\rmit{e.g.,}}              
\def\cf{cf.}                       
\def\viz{\rmit{viz.}}
\def\vs{\rmit{vs.}}

\def\rot{\hbox{\rm rot}}
\def\div{\hbox{\rm div}}
\def\lesssim{\mathrel{\hbox{\rlap{\hbox{\lower4pt\hbox{$\sim$}}}\hbox{$<$}}}}
\def\gtrsim{\mathrel{\hbox{\rlap{\hbox{\lower4pt\hbox{$\sim$}}}\hbox{$>$}}}}
\def\dif{\: {\rm d}}                       
\def\ep{\:{\rm e}^}                        
\def\dash{\hbox{$\,-\,$}}                  
\def\is{\!=\!}                             

\def\starname#1#2{${#1}$\,{\rm {#2}}}  
\def\Teff{\hbox{$T_{\rm eff}$}}   

\def\kms{\hbox{km$\;$s$^{-1}$}}
\def\ms{\hbox{m$\;$s$^{-1}$}}
\def\Mxcm{\hbox{Mx\,cm$^{-2}$}}    

\def\Bapp{\hbox{$B_{\rm app}$}}    

\def\komega{($k, \omega$)}                 
\def\kf{($k_h,f$)}                         
\def\VminI{\hbox{$V\!\!-\!\!I$}}           
\def\IminI{\hbox{$I\!\!-\!\!I$}}           
\def\VminV{\hbox{$V\!\!-\!\!V$}}           
\def\Xt{\hbox{$X\!\!-\!t$}}                

\def\level #1 #2#3#4{$#1 \: ^{#2} \mbox{#3} ^{#4}$}   

\def\specchar#1{\uppercase{#1}}    
\def\AlI{\mbox{Al\,\specchar{i}}}  
\def\BI{\mbox{B\,\specchar{i}}} 
\def\BII{\mbox{B\,\specchar{ii}}}  
\def\BaI{\mbox{Ba\,\specchar{i}}}  
\def\BaII{\mbox{Ba\,\specchar{ii}}} 
\def\CI{\mbox{C\,\specchar{i}}} 
\def\CII{\mbox{C\,\specchar{ii}}} 
\def\CIII{\mbox{C\,\specchar{iii}}} 
\def\CIV{\mbox{C\,\specchar{iv}}} 
\def\CaI{\mbox{Ca\,\specchar{i}}} 
\def\CaII{\mbox{Ca\,\specchar{ii}}} 
\def\CaIII{\mbox{Ca\,\specchar{iii}}} 
\def\CoI{\mbox{Co\,\specchar{i}}} 
\def\CrI{\mbox{Cr\,\specchar{i}}} 
\def\CriI{\mbox{Cr\,\specchar{ii}}} 
\def\CsI{\mbox{Cs\,\specchar{i}}} 
\def\CsII{\mbox{Cs\,\specchar{ii}}} 
\def\CuI{\mbox{Cu\,\specchar{i}}} 
\def\FeI{\mbox{Fe\,\specchar{i}}} 
\def\FeII{\mbox{Fe\,\specchar{ii}}} 
\def\FeIX{\mbox{Fe\,\specchar{ix}}}
\def\FeX{\mbox{Fe\,\specchar{x}}}
\def\FeXVI{\mbox{Fe\,\specchar{xvi}}}
\def\FrI{\mbox{Fr\,\specchar{i}}}
\def\HI{\mbox{H\,\specchar{i}}} 
\def\HII{\mbox{H\,\specchar{ii}}} 
\def\Hmin{\hbox{\rmH$^{^{_{\scriptstyle -}}}$}}      
\def\Hemin{\hbox{{\rm He}$^{^{_{\scriptstyle -}}}$}} 
\def\HeI{\mbox{He\,\specchar{i}}} 
\def\HeII{\mbox{He\,\specchar{ii}}} 
\def\HeIII{\mbox{He\,\specchar{iii}}} 
\def\KI{\mbox{K\,\specchar{i}}} 
\def\KII{\mbox{K\,\specchar{ii}}} 
\def\KIII{\mbox{K\,\specchar{iii}}} 
\def\LiI{\mbox{Li\,\specchar{i}}} 
\def\LiII{\mbox{Li\,\specchar{ii}}} 
\def\LiIII{\mbox{Li\,\specchar{iii}}} 
\def\MgI{\mbox{Mg\,\specchar{i}}} 
\def\MgII{\mbox{Mg\,\specchar{ii}}} 
\def\MgIII{\mbox{Mg\,\specchar{iii}}} 
\def\MnI{\mbox{Mn\,\specchar{i}}} 
\def\NI{\mbox{N\,\specchar{i}}}
\def\NIV{\mbox{N\,\specchar{iv}}}
\def\NaI{\mbox{Na\,\specchar{i}}}
\def\NaII{\mbox{Na\,\specchar{ii}}}
\def\NaIII{\mbox{Na\,\specchar{iii}}}
\def\NeVIII{\mbox{Ne\,\specchar{viii}}} 
\def\NiI{\mbox{Ni\,\specchar{i}}} 
\def\NiII{\mbox{Ni\,\specchar{ii}}}
\def\NiIII{\mbox{Ni\,\specchar{iii}}} 
\def\OI{\mbox{O\,\specchar{i}}} 
\def\OVI{\mbox{O\,\specchar{vi}}}
\def\RbI{\mbox{Rb\,\specchar{i}}} 
\def\SII{\mbox{S\,\specchar{ii}}} 
\def\SiI{\mbox{Si\,\specchar{i}}} 
\def\SiII{\mbox{Si\,\specchar{ii}}} 
\def\SrI{\mbox{Sr\,\specchar{i}}}
\def\SrII{\mbox{Sr\,\specchar{ii}}}
\def\TiI{\mbox{Ti\,\specchar{i}}} 
\def\TiII{\mbox{Ti\,\specchar{ii}}} 
\def\TiIII{\mbox{Ti\,\specchar{iii}}} 
\def\TiIV{\mbox{Ti\,\specchar{iv}}} 
\def\VI{\mbox{V\,\specchar{i}}} 
\def\HtwoO{\mbox{H$_2$O}}        
\def\Otwo{\mbox{O$_2$}}          

\def\Halpha{\mbox{H\hspace{0.1ex}$\alpha$}} 
\def\Ha{\mbox{H\hspace{0.2ex}$\alpha$}}
\def\Hbeta{\mbox{H\hspace{0.2ex}$\beta$}}
\def\Hgamma{\mbox{H\hspace{0.2ex}$\gamma$}}
\def\Hdelta{\mbox{H\hspace{0.2ex}$\delta$}}
\def\Hepsilon{\mbox{H\hspace{0.2ex}$\epsilon$}}
\def\Hzeta{\mbox{H\hspace{0.2ex}$\zeta$}}
\def\Lyalpha{\mbox{Ly$\hspace{0.2ex}\alpha$}}
\def\Lybeta{\mbox{Ly$\hspace{0.2ex}\beta$}}
\def\Lygamma{\mbox{Ly$\hspace{0.2ex}\gamma$}}
\def\Lycont{\mbox{Ly\hspace{0.2ex}{\small cont}}}
\def\Baalpha{\mbox{Ba$\hspace{0.2ex}\alpha$}}
\def\Babeta{\mbox{Ba$\hspace{0.2ex}\beta$}}
\def\Bacont{\mbox{Ba\hspace{0.2ex}{\small cont}}}
\def\Paalpha{\mbox{Pa$\hspace{0.2ex}\alpha$}}
\def\Bralpha{\mbox{Br$\hspace{0.2ex}\alpha$}}

\def\NaD{\mbox{Na\,\specchar{i}\,D}}    
\def\NaDone{\mbox{Na\,\specchar{i}\,\,D$_1$}}
\def\NaDtwo{\mbox{Na\,\specchar{i}\,\,D$_2$}}
\def\NaID{\mbox{Na\,\specchar{i}\,\,D}}
\def\NaIDone{\mbox{Na\,\specchar{i}\,\,D$_1$}}
\def\NaIDtwo{\mbox{Na\,\specchar{i}\,\,D$_2$}}
\def\Done{\mbox{D$_1$}}
\def\Dtwo{\mbox{D$_2$}}

\def\Mgbone{\mbox{Mg\,\specchar{i}\,b$_1$}}
\def\Mgbtwo{\mbox{Mg\,\specchar{i}\,b$_2$}}
\def\Mgbthree{\mbox{Mg\,\specchar{i}\,b$_3$}}
\def\MgIb{\mbox{Mg\,\specchar{i}\,b}}
\def\MgIbone{\mbox{Mg\,\specchar{i}\,b$_1$}}
\def\MgIbtwo{\mbox{Mg\,\specchar{i}\,b$_2$}}
\def\MgIbthree{\mbox{Mg\,\specchar{i}\,b$_3$}}

\def\CaIIK{\mbox{Ca\,\specchar{ii}\,K}}       
\def\CaIIH{\mbox{Ca\,\specchar{ii}\,H}}
\def\CaIIHK{\mbox{Ca\,\specchar{ii}\,H\,\&\,K}}
\def\HK{\mbox{H\,\&\,K}}
\def\Kthree{\mbox{K$_3$}}      
\def\Hthree{\mbox{H$_3$}}
\def\Ktwo{\mbox{K$_2$}}
\def\Htwo{\mbox{H$_2$}}
\def\Kone{\mbox{K$_1$}}     
\def\Hone{\mbox{H$_1$}}     
\def\KtwoV{\mbox{K$_{2V}$}}
\def\KtwoR{\mbox{K$_{2R}$}}
\def\KoneV{\mbox{K$_{1V}$}}
\def\KoneR{\mbox{K$_{1R}$}}
\def\HtwoV{\mbox{H$_{2V}$}}
\def\HtwoR{\mbox{H$_{2R}$}}
\def\HoneV{\mbox{H$_{1V}$}}
\def\HoneR{\mbox{H$_{1R}$}}

\def\hk{\mbox{h\,\&\,k}}
\def\kthree{\mbox{k$_3$}}    
\def\hthree{\mbox{h$_3$}}
\def\ktwo{\mbox{k$_2$}}
\def\htwo{\mbox{h$_2$}}
\def\kone{\mbox{k$_1$}}     
\def\hone{\mbox{h$_1$}}     
\def\ktwoV{\mbox{k$_{2V}$}}
\def\ktwoR{\mbox{k$_{2R}$}}
\def\koneV{\mbox{k$_{1V}$}}
\def\koneR{\mbox{k$_{1R}$}}
\def\htwoV{\mbox{h$_{2V}$}}
\def\htwoR{\mbox{h$_{2R}$}}
\def\honeV{\mbox{h$_{1V}$}}
\def\honeR{\mbox{h$_{1R}$}}

\ifnum\preprintheader=1     
\makeatletter  
\def\@maketitle{\newpage
\markboth{}{}%
  {\em \footnotesize To appear in ``Magnetic Coupling between the Interior 
       and the Atmosphere of the Sun'', eds. S.~S.~Hasan and R.~J.~Rutten, 
       Astrophysics and Space Science Proceedings, Springer-Verlag, 
       Heidelberg, Berlin, 2009.}\par
 \def\lastand{\ifnum\value{@inst}=2\relax
                 \unskip{} \andname\
              \else
                 \unskip \lastandname\
              \fi}%
 \def\and{\stepcounter{@auth}\relax
          \ifnum\value{@auth}=\value{@inst}%
             \lastand
          \else
             \unskip,
          \fi}%
  \raggedright
 {\Large \bfseries\boldmath
  \pretolerance=10000
  \let\\=\newline
  \raggedright
  \hyphenpenalty \@M
  \interlinepenalty \@M
  \if@numart
     \chap@hangfrom{}
  \else
     \chap@hangfrom{\thechapter\thechapterend\hskip\betweenumberspace}
  \fi
  \ignorespaces
  \@title \par}\vskip .8cm
\if!\@subtitle!\else {\large \bfseries\boldmath
  \vskip -.65cm
  \pretolerance=10000
  \@subtitle \par}\vskip .8cm\fi
 \setbox0=\vbox{\setcounter{@auth}{1}\def\and{\stepcounter{@auth}}%
 \def\thanks##1{}\@author}%
 \global\value{@inst}=\value{@auth}%
 \global\value{auco}=\value{@auth}%
 \setcounter{@auth}{1}%
{\lineskip .5em
\noindent\ignorespaces
\@author\vskip.35cm}
 {\small\institutename\par}
 \ifdim\pagetotal>157\p@
     \vskip 11\p@
 \else
     \@tempdima=168\p@\advance\@tempdima by-\pagetotal
     \vskip\@tempdima
 \fi
}
\makeatother     
\fi


\title*{A Topology for the Penumbral Magnetic Fields}


    \author{J.~S\'anchez~Almeida}

    \authorindex{S\'anchez~Almeida, J.}


   \institute{Instituto de Astrof\'\i sica de Canarias, 
              La Laguna, Tenerife, Spain}

\authorrunning{S\'anchez~Almeida}  

\maketitle

\setcounter{footnote}{0}  

\begin{abstract} 
	We describe a scenario for the  topology of 
the magnetic field in penumbrae that accounts for
recent observations showing upflows, downflows, and 
reverse magnetic polarities. According to our conjecture,
short narrow magnetic loops fill the penumbral 
photosphere. 
Flows along these arched field lines are responsible
for both the Evershed effect and the convective
transport.
This scenario seems to be 
qualitatively consistent with most existing observations,
including the dark cores in penumbral filaments
reported by Scharmer et al.\ Each  bright filament with
dark core would be a system of two paired convective rolls 
with the dark core tracing the 
common lane where the plasma  sinks down.
The magnetic loops would have a hot footpoint
in one of the bright filament and a cold 
footpoint in the dark core.
The scenario  fits in most of our theoretical
prejudices (siphon flows along field lines, presence
of overturning convection, drag of field lines by downdrafts,
etc). 
If the conjecture turns out to be  correct, 
the mild upward and
downward velocities observed in penumbrae must
increase upon 
improving the 
resolution. 
This and other observational
tests to support or
disprove the scenario are put forward.
\end{abstract}


\section{Introduction}\label{almeida-introduction}

We are celebrating the centenary 
of the discovery by John~\citet{almeida-eve09} of the 
effect now bearing his name. Photospheric spectral
lines in sunspots are systematically
shifted toward the red in the limb-side penumbra, and
toward the blue in the center-side penumbra. 
A hundred years have passed and, despite the remarkably
large number of works on the Evershed 
effect\footnote{The NASA Astrophysics
Data System provides more than fourteen hundred papers
under the keyword {\em penumbra}, seventy of them
published during the last year.\label{almeida-footnote1}}, 
we still ignore how and why these line shifts are 
produced 
\citep[see, e.g., the review paper by][]{almeida-tho04}. 
Thus, the Evershed effect is among the oldest
unsolved problems in astronomy. Although its study
has never disappeared from the specialized literature,
the Evershed effect has  undergone a recent revival triggered 
by the advent of new instrumentation 
\citep[][]{almeida-sch02,almeida-kos07}, original
theoretical ideas \citep{almeida-wei04,almeida-spr06}, as well as 
realistic numerical simulations \citep{almeida-hei07,almeida-rem08}. 
Unfortunately, this renewed interest has not 
come together with a renewal of the diagnostic techniques, 
i.e.,  the methods and procedures that allow us to infer
physical properties from observed images and 
spectra. Often implicitly, the observers assume 
the physical properties to be constant
in the resolution element, a working hypothesis
clearly at odds with the observations. Spectral line 
asymmetries show up even with our best spatial
resolution 
\citep[][\S~\ref{almeida-observations}]{almeida-ich07,almeida-san07b}.
This lack of enough resolution is not secondary. 
The nature of the
Evershed flow has remained elusive so far because we have been
unable to isolate and identify the physical processes 
responsible for the line shifts.
Different measurements provide different ill-defined
averages
of the same unresolved underlaying structure, thus 
preventing simple interpretations and
yielding the
problems of consistency that plague the 
Evershed literature (e.g., non-parallelism between 
magnetic field lines and flows, \citealt{almeida-are90};
violation of the conservation of magnetic flux, \citealt{almeida-san98a};
non-parallelism between continuum filaments and 
magnetic field lines, \citealt{almeida-kal91}).

Understanding the observed spectral line asymmetries
complicates our analysis but, in reward,
the asymmetries provide a unique
diagnostic tool. They arise
from sub-pixel variations of the magnetic fields and flows, 
therefore, by modeling and interpretation of asymmetries,
one can get a handle on the unresolved structure. 
Although indirectly, such modelling allows us to
surpass the limitations imposed by the finite resolution. 
The idea has tradition in penumbral research, 
starting from the discovery of the asymmetries 
almost fifty years ago \citep[e.g.,][]{almeida-bum60,almeida-gri72}.
\citet[][hereinafter SA05]{almeida-san04b} exploits the tool
in a systematic study that encompasses a full round
sunspot. The unresolved components
found by SA05 inspire the
topology for the penumbral magnetic fields proposed
here. According to SA05, the asymmetries of the Stokes 
profiles\footnote{We use Stokes parameters to characterize
the polarization; $I$ for the intensity,
$Q$ and $U$ for the two independent types
of linear polarization, and $V$ for the
circular polarization. The 
Stokes profiles are 
representations of $I$, $Q$, $U$ and $V$ versus 
wavelength for a particular spectral line.
They follow well defined symmetries when
the atmosphere has constant magnetic field 
and  velocity 
\citep[see, e.g.,][]{almeida-san96}.
}
can be {\em quantitatively} explained if magnetic fields 
having a polarity opposite to the sunspot main polarity 
are  common throughout the penumbra. The reverse polarity 
holds intense magnetic field aligned flows which, consequently, 
are directed downward. Counter-intuitive as it may be, 
the presence of such ubiquitous strongly redshifted 
reverse polarity has been directly observed with the
satellite HINODE
\citep{almeida-ich07}. This new finding supports the original
SA05 results, providing credibility to the constraints 
that they impose on the magnetic fields and mass flows.
The existence of such ubiquitous return of magnetic
flux, together with a number of selected
results from the literature, are 
assembled here to offer a
plausible scenario for the penumbral magnetic field 
topology. Such exercise  to piece together 
and synthesize information from different sources is 
confessedly speculative.
It will not lead to a self-consistent MHD model for 
the penumbral structure. However, the exercise is
worthwhile for a number of reasons.
First, the compilation of observational results and
theoretical arguments in \S~\ref{almeida-observations} 
provides a unique 
brief-yet-restrictive list of constraints to be
satisfied by any explanation of the Evershed 
effect deserving such name. This reference
compilation 
will be useful even if the proposed magnetic
topology turns out to be incorrect. 
Second, we will show how the presence of a 
reverse polarity is qualitatively consistent with 
all existing observations, including unobvious cases.
Third, despite the amount of observational
and theoretical papers on
penumbrae, our understanding of the Evershed 
phenomenon is far from satisfactory. The solution
to the riddle 
cannot be trivial, and it may require interpreting the
observations with an alternative twist. 
Unconventional proposals are needed, and 
here we explore some of the possibilities.
Finally, the proposal may help inspiring 
MHD numerical modelers.

The work is structured as follows: it
begins by summarizing a number of observations
to constrain the topology 
of the magnetic field and velocity in penumbrae
(\S~\ref{almeida-observations}). These observational 
results leave an important loose end. The reverse polarity found by \citet{almeida-ich07} 
does not show up in high spatial resolution observations
with the Swedish Solar Telescope \citep[SST;][]{almeida-lan05,almeida-lan07}.
\S~\ref{almeida-what is new} explains how the 
apparent inconsistency goes away 
if the dark cores in penumbral filaments
found by \citet{almeida-sch02} 
correspond to the strongly redshifted  
reverse polarity regions spotted by HINODE. 
Once the difficulty has
been cleared up, 
the actual scenario for the magnetic field topology
is put forward in \S~\ref{almeida-scenario}.
Its predictions are qualitatively 
compared with observations in \S~\ref{almeida-qua_obs}.
Similarities and differences between
this scenario and existing models for the
penumbral structure and the Evershed flow are
analyzed in \S~\ref{almeida-conclusions},
where we also  suggest observational 
tests to confirm
or disprove our conjecture.

%
%
\section{Constraints on the penumbral structure}
\label{almeida-observations}
	
	The bibliography on the magnetic structure
of penumbrae is too extensive to be condensed in a few
pages (see footnote~\#\,\ref{almeida-footnote1}).
We refrain from giving an overview, 
and those readers interested in a more formal
review should refer to, e.g.,
\citet{almeida-sch91},
\citet{almeida-tho92},
\citet{almeida-sol03},
\citet{almeida-bel04},
or \citet{almeida-tho04}. 
Only the selected references that provide 
a framework for our proposal are
introduced and discussed here. They include 
SA05 and \citet{almeida-ich07}. 
Most of them are pure observational results, but 
several items
involve theoretical arguments as well.
The selection is obviously biased in the sense 
that some observations often bypassed are emphasized
here, and vice versa. However, 
to the best of our knowledge,
no potentially important constraint has been excluded.

\begin{enumerate}
\item \label{almeida-best}
The best penumbral images have a resolution
of the order 0\farcs 12 (or 90~km on the Sun).
They show many features at the
resolution limit implying the existence
of unresolved substructure. For example,
the power spectrum of the penumbral images has
signal down to the instrumental cutoff \citep{almeida-rou04},
and the width of the narrower penumbral filaments is
set by the resolution of the observation 
(\citealt{almeida-sch02}; see also Fig.~\ref{almeida-coresfig}).
This interpretation of the current observations
should not be misunderstood. The penumbrae have
structures of all sizes starting with the penumbra
as a whole. However, the  observations show that
much of its observed structure is at the resolution
set by the present technical limitations and,
therefore, it is expected to be unresolved.
This impression is corroborated by the presence of  
spectral line asymmetries as discussed in 
item~\ref{almeida-ncp}.

\item \label{almeida-cores}
The best penumbral images show 
{\em dark cores in penumbral filaments}
\citep{almeida-sch02}. 
We prefer to describe them as  dark filaments
outlined by bright plasma.
This description also provides a fair account
of the actual observation (Fig.~\ref{almeida-coresfig}), 
but it 
emphasizes the role of the dark core. Actually,
dark cores without a bright side are 
common, and the cores seldom emanate from
a bright point (Fig.~\ref{almeida-coresfig}).
\begin{figure}
\centering
\includegraphics[width=1.0\textwidth]{\figspath/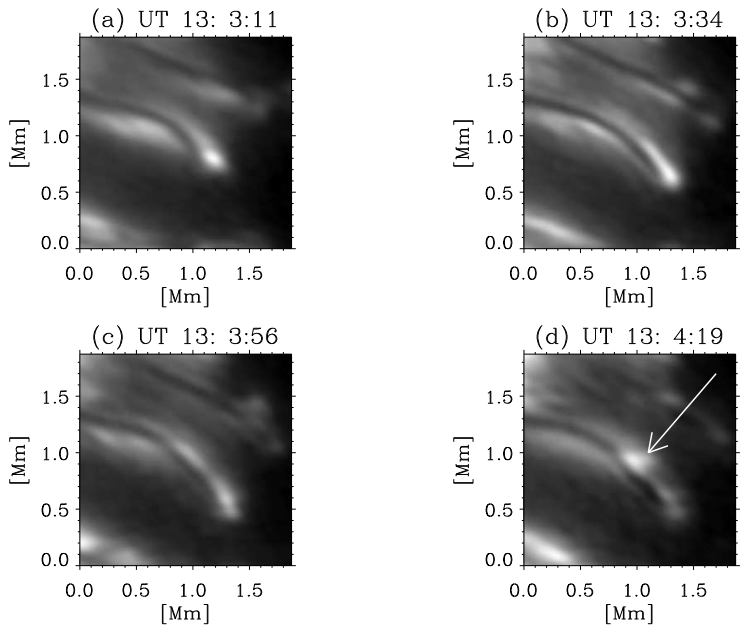}
\caption{Time evolution
of one of the {\em dark cores in penumbral filaments}
discovered by \citet{almeida-sch02}. (The
UT of observation is marked on top of each snapshot.)
Note that one of the
bright sides is partly missing  in (c) and (d). Note also
that the bright points are not on the dark filament
but in a side. These two
properties are common. The arrow indicates the
emergence of a new bright point in a side of
the pre-existing dark filament.
Note the narrowness of the bright filaments, and their
large aspect ratio (length over width).
The spatial scales are in Mm, and the 
angular resolution of the image is of the order
of 0.09~Mm.}
\label{almeida-coresfig}
\end{figure}
The widths of the dark core and its bright 
boundaries
remain  unresolved,
although the set formed by a dark core 
sandwiched between two bright filaments
spans some 150-180~km across.
\item \label{almeida-correlation}
	There is a local correlation 
	between penumbral brightness
	and Doppler shift, so that bright features
	are blueshifted with respect to  dark features 
	\citep{almeida-bec69c,almeida-san93b,almeida-san07b,almeida-joh93,almeida-sch00b}. 
	The correlation maintains the same sign 
	in the limb-side
	penumbra and the center-side penumbra, 
	a property invoked by \citet{almeida-bec69c} to conclude
	that it is produced by vertical motions. 
	A positive correlation between vertical velocity
	and intensity is characteristic of the 
	non-magnetic granulation.
	The fact that the same correlation
	also exists in penumbrae suggests
	a common origin for the two phenomena, namely,
	convection.
\item\label{almeida-this}
    The limb-side  and  center-side parts of a penumbra are slightly
    darker than the rest, an observational fact 
    indicating that the bright penumbral filaments are elevated with 
    respect to the dark ones \citep{almeida-sch04}. The behavior seems 
    to continue down to the smallest structures. Dark cores
    are best seeing where the low resolution penumbra is darkest 
    according to \citet{almeida-sch04}, i.e., along the center-to-limb 
    direction \citep[e.g.,][]{almeida-lan07,almeida-ich07b}. 
    The two observations are probably connected, suggesting that
    dark cores are depressed with respect to their bright
    sides.
\item There is a local correlation between 
	magnetic field inclination and horizontal velocity.
	The largest velocities are associated with
	the more horizontal fields
	\citep[e.g.,][]{almeida-tit93,almeida-sta97}.

\item \label{almeida-horizontal}
	The large horizontal motions occur in the 
	dark penumbral filaments \citep[e.g.,][]{almeida-rue99,almeida-pen03,almeida-san07b}.
	This trend continues down to the dark cores in penumbral 
	filaments \citep{almeida-lan05,almeida-lan07}.

\item \label{almeida-contradict}
	The observations on the correlation
between magnetic field strength and brightness
are contradictory. Some authors find the 
strongest field strengths associated with the
darkest regions, and vice versa (c.f. \citealt{almeida-bec69c} 
and \citealt{almeida-hof94}).
What seems to be clear is the reduced circular polarization 
signal existing in dark cores, which is commonly
interpreted as a  reduced  
field strength \citep{almeida-lan05,almeida-lan07}.
We show in \S~\ref{almeida-what is new} that such dimming
of the circular polarization admits a totally 
different interpretation, consistent with an increase
of field strength in dark cores.

\item	\label{almeida-roll2}
	Theoretical arguments indicate that
	the convective roll pattern should 
	be the  mode of convection 
	for nearly horizontal magnetic fields 
	\citep{almeida-dan61,almeida-hur00}. The rolls have their axes
	along the magnetic field lines.
	Unfortunately, this is not what results 
	from recent numerical simulations of magneto-convection
	in strong highly inclined magnetic fields 
	\citep{almeida-hei07,almeida-rem08}. Here the convection takes 
	place as field-free plasma intrusions in a strong field 
	background, resembling the gappy penumbra model
	by \citet{almeida-spr06}. However, 
	these numerical simulations many not be realistic enough. 
	They are the first to come in a series trying to
	reduce the artificial diffusivities employed by the 
	numerical schemes. It is unclear whether such evolution 
	will maintain the modes of convection.
	Moreover, even the present simulations hint at 
	the existence of a convective roll pattern.

\item  \label{almeida-roll1}
	The stable sunspots are surrounded by a large
	annular convection cell called moat \citep{almeida-she72}.
	The moat presents an outflow which, contrarily to the 
	commonly held opinion,
	has  both radial and tangential velocities  
	(Title 2003, private communication;
	\citealt{almeida-bon04}). The tangential component
	is well organized so that it sweeps the plasma 
	toward radial channels, creating a
	velocity pattern that resembles the convective
	rolls by \citet{almeida-dan61}. Compare Fig.~7
	in \citet{almeida-bon04} or Fig.~11 in
	\citet{almeida-bov03} with the rolls in Fig.~9a of  \citet{almeida-gar87}.

\item \label{almeida-drag}
	The magnetic field lines can be dragged by the 
	downdrafts of the granulation. 
	Termed as {\em flux pumping} mechanism,
	this drag is modeled and studied by \citet{almeida-wei04}
	and \citet{almeida-bru08}
	to show that the vigorous sinking plumes of the
	granulation and mesogranulation easily pumps down 
	magnetic fluxtubes outside the penumbra.
	It is conceivable that 
	the same pumping by sinking 
	plasma also operates
	within the magnetized penumbra. 

\item \label{almeida-ncp}
	The polarization  of the spectral lines emerging from 
	any sunspot has asymmetries and, consequently, it requires several 
	unresolved velocities and magnetic fields to be produced 
	\citep{almeida-bum60,almeida-gri72,almeida-gol74,almeida-san92b}. The asymmetries
	show up even with the best spatial resolution achieved
	nowadays (150~km -- 250~km; \citealt{almeida-san07b}, \citealt{almeida-ich07}).
	Furthermore, part of the 
	substructure producing asymmetries will never be resolved
	directly because it occurs along the line-of-sight (LOS).
	The spectral lines create 
	net circular polarization (NCP), i.e., the asymmetry
	of the Stokes~$V$ profiles is such that the
	integrated circular polarization of any typical spectral
	line is not zero. NCP can only be produced by gradients
	along the LOS and, therefore, within a range
	of heights smaller that the region where
	the lines are formed (say, 100~km).
	The NCP follows several general rules found by
	\citet{almeida-ill74a,almeida-ill74b} and \citet{almeida-mak86}. 
	Explaining them requires large gradients 
	of magnetic field inclination along the LOS \citep{almeida-san92b}.
	The gradients can be produced by the  
	discontinuities occurring at the boundary of 
	relatively large magnetic fluxtubes 
	embedded in a background \citep{almeida-sol93b}, 
	or as the cumulative effect of several smaller optically thin 
	structures with varied magnetic 
	field inclinations \citep{almeida-san96}.
\item \label{almeida-up_down} 
The sunspots seem to show upward motions in the inner penumbra and
downward motions in the outer penumbra
\citep[e.g.,][]{almeida-rim95,almeida-sch00,almeida-bel03b,almeida-tri04}.
However,
this velocity pattern is inferred 
assuming uniform velocities in the resolution elements,
an hypothesis inconsistent with  
observations (items~\ref{almeida-best} and \ref{almeida-ncp}).
\item \label{almeida-myroll}
Using a MISMA\footnote{
The acronym MISMA stands for 
MIcro-Structured Magnetic Atmosphere, and it
was coined by \citet{almeida-san96} to describe 
magnetic atmospheres having optically-thin
substructure.} framework,
SA05 carries out a systematic fit of 
all Stokes profiles 
observed with 1\arcsec-resolution in a medium sized sunspot. 
Line asymmetries and NCP are reproduced
(item~\ref{almeida-ncp}).
The resulting semi-empirical model sunspot 
provides both
the large scale magnetic structure, as well as the
small scale properties of the micro-structure.
On top of a regular large scale behavior,
the inferred small scale structure of the magnetic 
fields and flows is novel and unexpected. 
Some 30\% of the volume is occupied by
magnetic field lines that return to the 
sub-photosphere within the penumbral boundary. 
Mass flows are aligned with magnetic field lines, 
therefore, the field lines with the main sunspot
polarity transport mass upward, whereas the
reverse polarity is associated
with high speed flows returning to the
solar interior. 
This return of magnetic
flux and mass toward the solar interior occurs 
throughout the penumbra, as opposed to previous 
claims of bending over and return  at the penumbral 
border or beyond (item~\ref{almeida-up_down}).
The observed magnetic field strength 
difference between field lines pointing
up and down can drive a siphon flow with the magnitude and
sense of the Evershed flow.
Within observational uncertainties, the mass transported
upward is identical to the mass going downward. 
\item \label{almeida-cooling}
The bright penumbral filaments are too long to trace individual 
streams of hot plasma. The original
argument dates back to \citet{almeida-dan60}, but here we recreate a  
recent account by 
 \citet{almeida-sch99}. They estimate the length of a bright
filament produced by hot plasma flowing along a 
magnetic fluxtube. The plasma cools down as it radiates away and so, 
eventually, the fluxtube becomes dark
and transparent.
An isolated loop  would have 
a bright head whose length $l$ is approximately set  by 
the cooling time of the emerging plasma $t_c$ times the velocity
of the mass flow along the field lines $U$,
\begin{equation}
l\approx U t_c.
\end{equation}
The cooling time depends on the diameter of the tube $d$,
so that the thinner the tube the  faster the cooling.
For reasonable values of the Evershed flow speed 
(U$\approx$ 5~km~s$^{-1}$), and using the cooling time 
worked out by \citet{almeida-sch99}, the aspect ratio
of the hot footpoint turns out to be of the order
of one for a wide range of fluxtube diameters, i.e., 
\begin{equation}
l/d\approx 0.8~ (d/200~{\rm km})^{0.5}.
\end{equation}
Filaments must have $l/d >> 1$, and so, 
a hot plasma stream will show up as a bright knot rather
than as a filament. In other words, the cooling of hot plasma
moving along field lines cannot give rise
to the kind of observed  filaments
(see Fig.~\ref{almeida-coresfig}).
If arrays of hot plasma streams form the filaments, they
must be arranged with their
hot and cold  footpoints aligned to give rise to
the observed structures.
\item \label{almeida-ichimoto}
HINODE magnetograms of penumbrae obtained
in the far wings of 
Fe~{\sc i}~$\lambda$6302.5~\AA\
show a 
redshifted magnetic component with a polarity 
opposite to the main sunspot polarity 
\citep[Fig. 4 in][]{almeida-ich07}.
The patches of opposite polarity are 
scattered throughout the penumbra. 
In addition, this reverse polarity is 
associated with extremely asymmetric 
Stokes~$V$ profiles having three lobes
\citep[][Fig. 5, and Ichimoto 2008, 
private communication]{almeida-ich07}.
These asymmetric profiles are known in the 
traditional literature as cross-over effect,
a term used along the paper.

\end{enumerate}

\section{Model MISMA for penumbral filaments with
	dark core}\label{almeida-what is new}

The set of constraints summarized in \S~\ref{almeida-observations}
leave an important loose end. The reverse polarity predicted
by SA05 (item~\ref{almeida-myroll}) and found by 
\citet[][item~\ref{almeida-ichimoto}]{almeida-ich07} 
does not show up in SST magnetograms 
\citep[][and \S~\ref{almeida-introduction}]{almeida-lan05,almeida-lan07}. 
This fact poses a serious problem since SST has 
twice HINODE spatial resolution and, therefore,
it should be simpler for SST to detect 
mixed polarities. 
Here we  show how HINODE and SST observations 
can be naturally understood within the two 
component model MISMA  by SA05,
provided that the dark cores are associated 
with an enhanced contribution of the reverse polarity.

\begin{figure}
\centering
\includegraphics[width=1.\textwidth]{\figspath/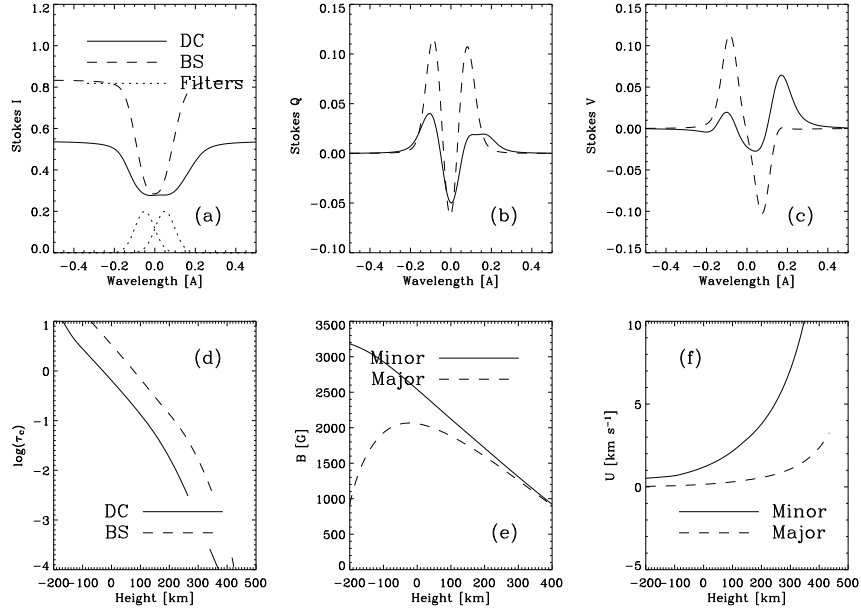}
\caption{
(a) Stokes~$I$ profiles in one of the representative
model MISMAs in SA05, which has been
slightly modified to represent a dark core 
(the solid line), and its bright sides (the dashed line).
They are normalized to the quiet Sun continuum intensity.
(b) Stokes~$Q$ profiles. 
(c) Stokes~$V$ profiles.
(d) Continuum optical depth $\tau_c$ vs height in the
	atmosphere for the dark core and the bright
sides, as indicated in the inset.
(e) Magnetic field strength vs height for the two
magnetic components of
the model MISMA. They are identical for 
the dark core and  the bright sides.
(f) Velocities along the magnetic field lines for
the two magnetic components of the model MISMA.
They are identical for 
the dark core and the bright sides. 
}
\label{almeida-hinode0}
\end{figure}

According to SA05, the component that contains most of the mass
in each resolution element has the polarity of the sunspot, and
bears mild magnetic-field-aligned mass flows.  
It is usually combined with a minor component
of opposite polarity and having large velocities.
In 1\arcsec-resolution observations,  the outcoming
light is systematically dominated by the major component, and 
the resulting Stokes profiles have rather regular 
shapes. An exception occurs in the so-called 
{\em apparent neutral line}, where the 
cross-over effect was discovered
\citep[see][and references therein]{almeida-san92b}. 
At the neutral line the mean magnetic 
field vector is perpendicular to the LOS, and 
the contribution  of the major component almost 
disappears due to projection effects. 
The improvement of angular resolution
augments the chances of finding the
minor component in excess, allowing
the opposite polarity to show up.
In order to illustrate the effect, 
several randomly chosen model MISMAs in SA05 
were modified by increasing the fraction of 
atmosphere occupied by the minor component.
One example is shown in 
Fig.~\ref{almeida-hinode0}.  
The resulting Stokes~$I$, $Q$ and $V$
profiles 
of {Fe}~{\sc i}~$\lambda$6302.5~\AA\ are represented
as solid lines in Figs.~\ref{almeida-hinode0}a,
\ref{almeida-hinode0}b and \ref{almeida-hinode0}c, 
respectively. They model a point in the 
limb-side penumbra at $\mu=0.95$ (18\deg\ 
heliocentric angle). 
Note how Stokes~$I$ is redshifted and
deformed, and how  Stokes~$V$ shows the cross-over
effect. 
Consequently, the improvement of spatial resolution
with respect to traditional earth-based spectro-polarimetric
observations naturally explains the abundance 
of cross-over Stokes~$V$ profiles found by HINODE. 
Figures~\ref{almeida-hinode0}a,  \ref{almeida-hinode0}b,
and \ref{almeida-hinode0}c
also show the case where the major component
dominates (the dashed line). The strong
asymmetries have disappeared, rendering 
Stokes~$V$ with antisymmetric  shape and the 
sign of the dominant
polarity. The magnetic field vector and 
the flows  are identical
in the two sets of Stokes profiles 
(Figs.~\ref{almeida-hinode0}e and 
\ref{almeida-hinode0}f), although, we have globally 
decreased
the temperature of the atmosphere giving rise to
the asymmetric profiles to mimic 
dark features (see the Stokes~$I$ continua 
in Fig.~\ref{almeida-hinode0}a).

\begin{figure}
\centering
\includegraphics[width=0.8\textwidth]{\figspath/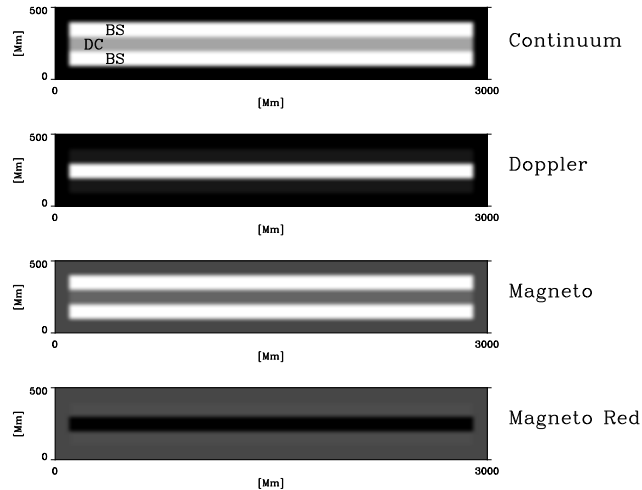}
\caption{
Schematic modeling of SST observations
of penumbral filaments by \citet{almeida-lan05,almeida-lan07}.
A dark core (DC) surrounded by two
bright sides (BS) 
is located in the limb-side 
penumbra of a sunspot at $\mu=0.95$
(18\deg\ heliocentric angle).
The three top images show a
continuum image, a dopplergram,
and a magnetogram, as labelled. 
The convention is such that both the 
sunspot main polarity
and a redshift produce positive signals.
The dark background in all
images has been included for reference, and
it represents signal equals zero. 
The fourth image ({\tt Magneto Red}) 
corresponds to a magnetogram in the far red wing 
of Fe~{\sc i}~$\lambda$6302.5~\AA , and it reveals a 
dark core with a polarity opposite
to the sunspot main polarity.
The continuum image and the dopplergram
have been scaled from zero (black) to 
maximum (white). The scaling of the two
magnetograms is the same, so that their 
signals can be compared directly.
}
\label{almeida-hinode1}
\end{figure}

Understanding HINODE observations in terms 
of MISMAs also explains the lack of reverse polarity
in SST magnetograms.  Stokes~$V$ in 
reverse polarity regions shows cross-over
effect 
\citep[Fig.~5 in ][ and Ichimoto 2008, 
private communication]{almeida-ich07},
i.e.,
it presents two polarities depending on the 
sampled wavelength. 
It has the main sunspot polarity
near line center, whereas the polarity 
is reversed in the far red wing
(see the solid line in Fig.~\ref{almeida-hinode0}c).
SST magnetograms are taken at line center 
($\pm 50$\ m\AA ), which explains why the 
reverse polarity does not show up.
A significant reduction of
the Stokes~$V$ signal occurs, though. Such reduction
automatically explains the observed
weakening of magnetic signals in 
dark cores (item~\ref{almeida-contradict} in 
\S~\ref{almeida-observations}), provided that
the dark cores are associated with an 
enhancement of the opposite polarity,
i.e., if the cross-over profiles are 
produced in the dark cores.  We have 
constructed images, magnetograms, and dopplergrams 
of a (na\"\i ve) model dark-cored filament 
that illustrate the idea. The filament
is formed by a uniform 100 km wide dark strip,
representing the dark core,  
bounded by two bright strips of the same width,
representing the bright sides.  
The Stokes profiles of the dark core have been 
taken as the solid lines in Figs.~\ref{almeida-hinode0}a 
and \ref{almeida-hinode0}c, whereas the bright sides are modelled
as the dashed lines in the same figures.
The color filters employed 
by \citet{almeida-lan05,almeida-lan07}
are approximated by Gaussian functions 
of 80\,m\AA\ FWHM, and shifted 
$\pm 50$\,m\AA\ from the line center 
(see the dotted lines in Fig.~\ref{almeida-hinode0}a).
The magnetogram signals are computed from the
profiles as  
\begin{equation}
M={\int V(\lambda) f(\lambda-\Delta\lambda)\,{\rm d}\lambda}\Big/
	\int I(\lambda) f(\lambda-\Delta\lambda)\,{\rm d}\lambda, 
\end{equation}
with $f(\lambda)$ the transmission curve of the
filter and $\Delta\lambda=-50$\,m\AA . 
Similarly, the Doppler signals are given by
\begin{equation}
D= \int I(\lambda) [f(\lambda-\Delta\lambda)-f(\lambda+\Delta\lambda)]\,
	{\rm d}\lambda\Big/\int I(\lambda) [f(\lambda-\Delta\lambda)+
	f(\lambda+\Delta\lambda)]\,{\rm d}\lambda,
\end{equation}
but here we employ the Stokes~$I$ profile of the non-magnetic
line used by 
\citeauthor{almeida-lan07}~(\citeyear{almeida-lan07}; 
i.e., {Fe}~{\sc i}~$\lambda$5576~\AA).
The signs of $M$ and $D$ ensure 
$M >0$ for the main polarity of the sunspot, 
and also $D> 0$ for redshifted profiles.
The continuum intensity 
has been taken as $I$ at -0.4\,\AA\ from the
line center. The continuum image of this model filament
is shown in Fig.~\ref{almeida-hinode1}, with the dark core 
and the bright sides marked as DC and BS, respectively.
The dopplergram and the magnetogram are also included in the same
figure. The dark background in all images indicates 
the level corresponding to no signal.
In agreement with \citeauthor{almeida-lan07} observations,
the filament shows redshifts ($D > 0$), which are 
enhanced in the dark core. 
In agreement with \citeauthor{almeida-lan07},
the filament shows the main polarity of the sunspot
($M > 0$), with the signal
strongly reduced in the dark core.
Figure~\ref{almeida-hinode1}, bottom, 
includes the 
magnetogram to be observed at the far red wing
($\Delta\lambda=200$\,m\AA). The dark core now shows
the reversed polarity ($M< 0$), whereas the bright 
sides still maintain the main polarity with an 
extremely weak signal. This specific
prediction of the modeling is liable for direct
observational test (\S~\ref{almeida-conclusions}).

Two final remarks are in order. First, the magnetogram signal
in the dark core is much weaker than  in the
bright sides, despite the fact that the (average)
magnetic field strength is larger in the core
(see Fig.~\ref{almeida-hinode0}e, keeping in mind that
the minor component dominates).
Second, the model dark core is depressed
with respect to the bright sides. 
Figure~\ref{almeida-hinode0}d shows the continuum 
optical depth $\tau_c$ as a function of the 
height in the atmosphere. When the two  atmospheres 
are in lateral pressure balance, 
the layer $\tau_c=1$ of the
dark core is shifted by some 100~km downward
with respect to the same layer in the 
bright sides. 
The depression of the observed
layers in the dark core is  produced by two effects; 
the decrease of density associated with the increase 
of magnetic pressure \citep[e.g.,][]{almeida-spr76}, 
and the decrease of opacity associated
with the reduction of temperature \citep[e.g.,][]{almeida-sti91}.

%
%
\section{Scenario for the 
	small-scale structure 
	of the penumbra}\label{almeida-scenario}

\begin{figure}
\centering
\includegraphics[width=0.8\textwidth]{\figspath/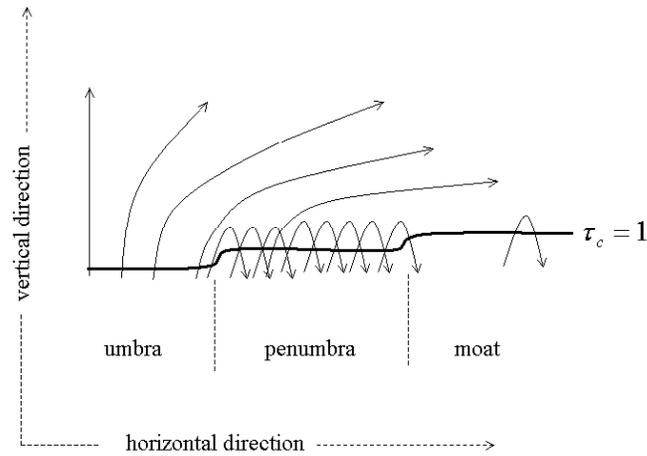}
\caption{Cartoon sketching the scenario for the penumbral
	magnetic field topology put forward in the paper.
	Magnetic field lines, represented as solid
	lines with arrow heads, return under the photosphere
	in the entire penumbra. The symbol $\tau_c$ stands
	for the continuum optical depth so that the thick solid line
	marked as $\tau_c=1$ represents the base of the photosphere.
	}
\label{almeida-cartoon2}
\end{figure}

\begin{figure}
\centering
\includegraphics[width=0.7\textwidth]{\figspath/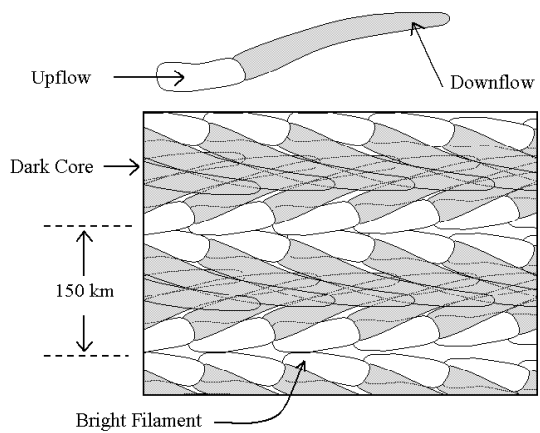}
\caption{A view from above of a small
	portion of penumbra (see the scale on the left hand side 
	of the cartoon).  Small magnetic loops like the one on top
	of the figure are averaged in our resolution element 
	(the rectangle).
	They are so thin that various loops overlap
	along the LOS.
	The loops are arranged with the downflowing footpoints
	aligned forming a dark core. The hot upflows
	feeding a dark core give rise to two bright filaments.
	The Evershed flow is directed along the field lines
	toward the right.}
\label{almeida-cartoon1}
\end{figure}

Attending to the constraints presented
in \S~\ref{almeida-observations},
penumbrae may be made out of short narrow shallow  magnetic 
loops which often return under the photosphere 
within the sunspot boundary (Fig.~\ref{almeida-cartoon2}). 
One of the footpoints is hotter than the other
(Fig.~\ref{almeida-cartoon1}).
The matter emerges in the hot footpoint, radiates away, cools down,
and returns through the cold footpoint.
The ascending plasma is hot,  dense, and  slowly moving.
The descending plasma is cold, tenuous, and  fast moving.
The motions along magnetic field lines
are driven by magnetic field strength
differences between the two footpoints, as
required by the siphon flow mechanism.

\begin{figure}
\centering
\includegraphics[width=0.8\textwidth]{\figspath/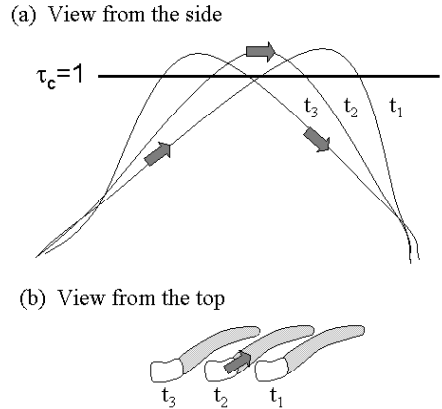}
\caption{Cartoon showing the time evolution 
	of a single magnetic field line, with $t_1 < t_2 < t_3$
	(the thin solid lines). 
	(a) View from the side, with the
	vertical direction pointing upward. The thick solid line
	corresponds to base of the photosphere (continuum optical 
	depth equals one). The Evershed flow is directed
	to the right.
	The thick arrows represent a parcel of fluid with the 
	tip pointing in the direction of the plasma motion.
	(b) View from above of the same three instants.
	(Compare with Fig.~\ref{almeida-cartoon1},
	showing many field lines at a given time.) It only shows
	the fluid parcel at $t_2$ since it remains below the 
	photosphere at $t_1$ and $t_3$.
	}
\label{almeida-cartoon3}
\end{figure}

In addition to holding large velocities along field lines,
the cold footpoint of each loop sinks down
in a slow motion across field lines. 
In non-magnetic convection, upflows are driven
through mass conservation by displacing
warm material around the downdrafts
\citep{almeida-stei98,almeida-ras03}. The uprising hot material tends
to emerge next to the downflows. 
If the same mechanism holds in penumbrae, the
sinking of cold footpoints induces a rise of the 
hot footpoints physically connected to them, 
producing  a backward 
displacement of the visible part of the loops 
(see Fig.~\ref{almeida-cartoon3}).  
The sink of the cold footpoints could be forced by the drag 
of  downdrafts in  subphotospheric
layers, in a magnetically modified 
version of the mechanism discussed 
in item~\ref{almeida-drag} of
\S~\ref{almeida-observations}.
The properties of the loops (length, width, speed,
and so on) should change with the 
position on the penumbra, but they are always 
narrow and so tending to be optically-thin
across field lines. One does not detect individual loops 
but assembles of them interleaved along the LOS.
The cold legs of many different loops should be
identified with the dark cores found by 
\citet{almeida-sch02};
compare Figs.~\ref{almeida-coresfig} and \ref{almeida-cartoon1}. 
The upflows of many loops account for be bright
penumbral filaments.
In this
scenario, magnetic field lines are not exactly aligned
with the penumbral bright and dark filaments, but the field lines
diverge from the bright filaments and converge
toward the dark filaments.
The mean field is radial, though.

According to our scenario,
bright penumbral filaments are associated with
fields having the polarity of the sunspot,
mild upflows, and relatively
low field strengths. On the contrary,
dark filaments are associated with 
fields whose polarity tends to be opposite to the
sunspot polarity, have intense flows, 
high field strength, and they are more transparent than 
the bright filaments.
%

%
\section{Qualitative comparison with observations}\label{almeida-qua_obs}
As we will discuss here, the scenario proposed 
in \S~\ref{almeida-scenario} fits in the  observations 
in \S~\ref{almeida-observations}. 
Our model penumbra is formed by narrow loops whose length is typically 
smaller than the penumbral size (Fig.~\ref{almeida-cartoon2}),
and whose width is not spatially resolved. The smallness
of the physical scales is in agreement
with items~\ref{almeida-best} and
\ref{almeida-ncp}. (These and the rest of numbers
refer to the labels in \S~\ref{almeida-observations}.) 
Magnetic field lines 
bend over and return under the photosphere over the entire penumbra, 
as required by 
items~\ref{almeida-myroll} and \ref{almeida-ichimoto}. 
The loops have a
hot footpoint with upward motion and a cold footpoint with 
downward motion, in agreement with the local correlation between
brightness and upward velocity observed 
in penumbrae
(item~\ref{almeida-correlation}). The downflows are expected to be faster 
than the upflows
since they are accelerated by the magnetic field strength difference
between the two footpoints, an image that  fits in
well  the observations showing the largest  
velocities to be associated with the dark penumbral
components (item~\ref{almeida-horizontal}).

We identify the  dark cores found
by \citet[][item~\ref{almeida-cores}]{almeida-sch02}
with cold footpoints of many loops, as 
sketched in Fig.~\ref{almeida-cartoon1}.
Dark cores trace downdrafts
engulfing cold footpoints (item~\ref{almeida-drag}).
The bright filaments
around the dark cores would be naturally explained 
by the presence of the downflows, as it happens 
with the enhanced brightness at the borders
of the granules in non-magnetic convection. 
Mass conservation induces an upflow
of hot material around the downdrafts 
\citep{almeida-ras95,almeida-ras03,almeida-stei98}.
The same mechanism
would produce the upraise of hot (magnetized) material
around the dark cores, forming two
bright filaments outlining each core
(item~\ref{almeida-cores}; Fig.~\ref{almeida-coresfig}). The hot
magnetized material would eventually cool down and
sink into the dark core to re-start the process.
In other words, a dark core would be the downdraft of 
two paired convective rolls, 
resembling those proposed long ago by Danielson 
(item~\ref{almeida-roll1}). In this
case, however, the magnetic field lines are not
exactly horizontal, and the plasma has a large
velocity component along the field
lines. Note that these hypothetical
convective rolls reproduce the expected mode of 
convective transport in highly inclined magnetic fields
(see item~\ref{almeida-roll2}, including the comment on the
recent numerical simulations of penumbrae which 
seem to disfavor this mode).
Moreover, a pattern of motions similar to these
convective rolls occurs in the moat surrounding
the sunspot  (item~\ref{almeida-roll1}), 
and it is conceivable that
it continues within the sunspot.
The existence of small scale convective 
upflows and downflows  does
not contradict the systematic
upward motions in the inner penumbra and
downward motions in the outer penumbra
found by various authors (see item~\ref{almeida-up_down}).
Most observational techniques employed so far 
assume uniform velocities in the resolution element.
When spatially unresolved 
upflows and downflows 
are interpreted as a single
resolved component, the measured
velocity corresponds to an ill-defined mean 
of the actual velocities. 
The contribution of upflows and downflows
to such mean is not proportional to the mass 
going up and down.
It depends on the physical
properties of the upflows and downflows, as well
as on the method employed to measure. The
mean vertical flux of mass inferred by SA05
is zero (item~\ref{almeida-myroll}), however the local
averages are biased\footnote{The effect is similar
to the convective blueshift of the spectral lines formed
in the granulation, whose existence does not imply a net
uplifting of the quiet photosphere.},
showing net upflows in the inner 
penumbra and net downflows in the outer penumbra, in agreement
with item~\ref{almeida-up_down}.

Our scenario with overlaying loops of various
velocities and inclinations accounts for the
observed Stokes asymmetries, including  
the rules for the NCP mentioned in item~\ref{almeida-ncp}.

The bright filaments are more opaque than the
dark cores (\S~\ref{almeida-what is new}), 
and they tend to block the light coming
from the dark cores when the filaments are 
observed sideways. This depression 
of the dark cores explains why they are elusive
in the penumbra perpendicular to the 
center-to-limb direction, as well as why
penumbrae are slightly darker in the line along the
center-to-limb direction (item~\ref{almeida-this}).

The length of the bright filaments is not set 
by the cooling time of individual fluxtubes, which
avoids the difficulty posed in item~\ref{almeida-cooling}.
It is given by the length of the dark core.

{\bf Does the model account for the penumbral radiative
flux?}\label{almeida-not_enough}
The radiative flux emanating from penumbrae $F$ is some
75\% of the flux in the quiet Sun. In order to
balance this loss with energy transported by 
convection,  the vertical velocity $U_z$ must satisfy
\citep[e.g.,][]{almeida-spr87,almeida-stei98}, 
\begin{equation}
U_z\approx F/(\rho\alpha \epsilon)\approx 1~{\rm km~s^{-1}},
\label{almeida-vert_veloc}
\end{equation}
with $\rho$ the density, $\alpha$ fraction
of atmospheric volume occupied by upward
motions, and $\epsilon$ the energy per unit mass
to be radiated away. Since the physical conditions
in penumbrae are similar to those of the quiet
Sun, the $U_z$ accounting for $F$
must be similar too, rendering the speed  in
the right-hand-side of 
equation~(\ref{almeida-vert_veloc}).
Unfortunately, the observed upward 
vertical velocities 
are one order of magnitude smaller
than the requirement set by  
equation~(\ref{almeida-vert_veloc})\footnote{This 
discrepancy  between the required and 
observed velocities  was used
to discard the transport of energy 
by convection in penumbrae 
\citep{almeida-spr87}, leading to the concept
of shallow penumbra by \citet{almeida-sch86b}.} 
(items~\ref{almeida-correlation} 
and \ref{almeida-myroll}).
The discrepancy can be explained if  
an observational bias 
underestimates the true velocities.  
Such bias is to be expected because the 
velocity structure remains unresolved
(items~\ref{almeida-best} and \ref{almeida-ncp}). 
Removing the
bias involves resolving the structure
both along and across the LOS, in particular,
the cross-over effect Stokes~$V$ profiles
associated with the reverse polarity must be properly
interpreted to retrieve realistic velocities. 
 
%
{\bf Why does the low-density plasma
of the cold footpoints sink rather than float?}
We have been arguing by analogy with the
non-magnetic convection, where the
(negative) buoyancy forces in the intergranular
lanes drive the sinking of cold plasma, and the
rising of hot material around it. The plasma 
tends to sink down due to its enhanced density
as compared to the hot upwelling plasma.
The scenario for the penumbral
convection discussed above does not reproduce
this particular aspect of the granular
convection. The  descending footpoint
has reduced density as compared to
the upflowing footpoint. The density in the
descending leg is lower than in the ascending leg,
and one may think that the descending plasma
is buoyant.
However, 
the density of the cold leg has to be compared to the
local density in the downdraft, which 
can easily be larger than the downdraft
density. Recall that the downdrafts have low
temperature and high magnetic field strengths,
two ingredients that naturally produce
low densities in 
magneto-hydrostatic equilibrium 
\citep[see, e.g.,][]{almeida-san01c}.

{\bf Why are the dark cores radially
oriented?}
According to our conjecture, the dark cores
are not tracing individual field lines. They are faults
in the global structure of the sunspot
where downflow motions are easier.
	 Such discontinuities of the global
magnetic structure would be favored 
if they are aligned with the mean magnetic
field, as it happens with  interchange 
instabilities 
\citep{almeida-par75,almeida-mey77,almeida-jah94}. 
The dark cores would  
be oriented along the direction of the 
mean penumbral magnetic field.

%

%

\section{Discussion and  tests}
\label{almeida-conclusions}

	We have described  a penumbral magnetic field
formed by short magnetic loops most of which
return to the sub-photosphere within
the sunspot boundary (Fig.~\ref{almeida-cartoon2}). 
Matter flowing along magnetic field lines would 
give rise to the Evershed effect. This flow along 
field lines would also be responsible for the
convective transport of heat in penumbrae.
Cold downflows of many loops are aligned in a sort
of lane of downdrafts whose observational
counterpart would be 
the dark cores in penumbral filaments found by
\citet{almeida-sch02}.
To some extent, the scenario
resembles the convective rolls put forward by 
Danielson some 50 years ago, except that (a) 
the mass also flows along field lines, and (b)
the mass may not re-emerge
after submergence. It is also
akin to the interchange convection 
\citep[e.g.,][]{almeida-sch02d}, where flows
along field lines transport heat from below
and give rise to the Evershed effect phenomena.
In our case, however,  the 
upraise of the hot tubes is induced
by the presence of downdrafts and the need to satisfy
mass conservation (as in the non-magnetic
granular convection). 
The submergence of the cool footpoints
of the loops happening in the downdrafts
may be due to flux pumping,
as numerical models have shown to
occur in the non-magnetic
downdrafts outside penumbrae \citep{almeida-wei04}.
Our scenario also owes some properties to the
siphon flow model \citep{almeida-mey68,almeida-tho93}. A gas
pressure difference between the loop footpoints drives
the  flows, but this difference is set by the 
magnetic field strength difference between
the hot and the cold penumbral structures,
rather than from field strength 
differences between 
the penumbra and magnetic concentrations
outside the sunspot.
Finally, the model that emerges has features from the return
flux model of \citet{almeida-osh82}, where the field lines
forming umbra and penumbra differ because the former
are open whereas the latter return to the photosphere.
The return of magnetic flux is included in our scenario, except that the
return of penumbral field lines occurs throughout
the penumbra, rather than outside the 
sunspot border.

This scenario seems to be compatible with most observations
existing in the literature, in particular, with 
the ubiquitous downflows and the reverse
magnetic flux found by SA05 and \citet{almeida-ich07} 
(see \S~\ref{almeida-qua_obs}). 
However, two difficulties remain. 
First, the upflows and downflows measured
so far are not fast enough to transport
the radiative flux  emerging
from penumbrae (\S~\ref{almeida-not_enough}).
Second, the reverse polarity does not show up
in SST observations, which nevertheless provide
the highest spatial resolution available at present 
(\S~\ref{almeida-what is new}). 
We believe that
the two difficulties are caused by
our still insufficient resolution. 
Solving the problem is not 
only a question of improving the resolution
across the LOS (e.g., by enlarging the telescope diameter), 
but the resolution along the LOS seems to be critical. 
The finding by \citet{almeida-ich07} that 
the reverse polarity coincides with 
very asymmetric Stokes profiles with NCP 
strongly suggests that measuring the true
velocities and polarities demands
resolution along the LOS. In other words,
the observed polarization cannot be correctly 
interpreted if one assumes the light to 
come from an atmosphere with uniform
properties.  

Several specific
tests will allow us to confirm or falsify 
the scenario.
(1) Diffraction limited high resolution spectro-polarimetry
with 1-meter class  telescopes  must
show a correlation between 
brightness and velocity with an amplitude of the 
order of 1~km~s$^{-1}$. As we point out above,
a pure brute force approach 
may not suffice to render the right amplitude, and
modeling the observed asymmetries seems to 
be required.
(2) SST magnetograms taken in the far 
red wing of Fe~{\sc i}~6302.5~\AA\ must show dark
cores with reverse polarity (\S~\ref{almeida-what is new}).
(3) The molecular spectra trace the cold 
penumbral component \citep[e.g.,][]{almeida-pen03}. They should show
a global downflow of the order of
1~km~s$^{-1}$, no matter the angular resolution
of the observation. 
(4) Local correlation tracking techniques
applied to sequences of penumbral images
should reveal proper motions of
bright features moving toward dark filaments. 
Such motions are masked
by the tendency of the bright plasma to cool down 
and become dark.
A global trend is to be expected, though.
Such trend seems to be present in high
resolution SST images \citep{almeida-mar05},
although the result requires independent 
confirmation.
\begin{acknowledgement}
In a paper for non-specialists, \citet{almeida-pri03}
speculates 
{\em ``Are the bright filaments doubly
convective rolls with a dark core that are cooling and
sinking?''} This question seems to advance
the scenario advocated here.
In addition,
the scenario is sketched in an unpublished work 
by \citet{almeida-san04d}, and it is also mentioned 
in \citet{almeida-san06}.
Using arguments differing from those introduced
here, \citet{almeida-zak08} have recently claimed evidence for
convective rolls in penumbrae. 
The image in Fig.~\ref{almeida-coresfig} is
courtesy of the Institute for Solar Physics
of the Royal Swedish Academy of Sciences, and it
was obtained with the SST operated in the Spanish Observatorio 
del Roque de Los Muchachos (La Palma). 
The work has partly been funded by the Spanish Ministry of Science
and Technology, project AYA2007-66502, as well as by
the EC SOLAIRE Network (MTRN-CT-2006-035484).
I thank the SOC of the Evershed centenary meeting
for this opportunity to expound unconventional
ideas.
\end{acknowledgement}

%

%
\begin{small}

\end{small}

\end{document}